\documentclass[twocolumn,aps,prb,unsortedaddress,superscriptaddress,showpacs]{revtex4}
\usepackage{graphicx}
\usepackage{dcolumn}
\usepackage{bm}
\usepackage{color}
\usepackage{ulem}

\begin{document}
\title{Self-consistent ac quantum transport using nonequilibrium Green functions}
\author{Diego Kienle}
\email[Present address: Theoretische Physik~I, Universit\"at Bayreuth, 95440 Bayreuth,
Germany; diego.kienle@uni-bayreuth.de]{}
\affiliation{Sandia National Laboratories, Livermore, California 94550, USA}
\author{Mani Vaidyanathan}
\affiliation{Department of Electrical and Computer Engineering, University of Alberta,
Edmonton, Alberta, Canada T6G 2V4}
\author{Fran\c{c}ois L\'{e}onard}
\date{\today}
\affiliation{Sandia National Laboratories, Livermore, California 94550, USA}

\begin{abstract}
We develop an approach for self-consistent ac quantum transport in the
presence of time-dependent potentials at non-transport terminals.
We apply the approach to calculate the high-frequency characteristics
of a nanotube transistor with the ac signal applied at the gate terminal.
We show that the self-consistent feedback between the ac charge and potential
is essential to properly capture the transport properties of the system.
In the on-state, this feedback leads to the excitation of plasmons, which
appear as pronounced divergent peaks in the dynamic conductance at terahertz
frequencies. In the off-state, these collective features vanish, and the
conductance exhibits smooth oscillations, a signature of single-particle
excitations. The proposed approach is general and will allow the study of the
high-frequency characteristics of many other low-dimensional nanoscale materials
such as nanowires and graphene-based systems, which are attractive for terahertz
devices, including those that exploit plasmonic excitations.
\end{abstract}

\pacs{72.10.Bg, 72.30.+q, 73.22.Lp, 73.63.Fg}

\maketitle

\section{INTRODUCTION}
A fundamental understanding of the physical processes controlling the
complex space- and time-dependent behavior of the carrier dynamics
in reduced dimensions is essential to assess the technological potential
of a variety of nanomaterials for future high-speed electronic devices.
Such assessment, however, requires both experimental and theoretical
techniques by which high-frequency material properties, such as the
dynamic (ac) conductance, can be measured or calculated.

Experimentally, much progress has been made in developing techniques to
probe the rf response of nanomaterials. Among these nanomaterials, carbon
nanotubes have received much attention due to their exceptional electronic
transport properties at dc,\cite{DurkopNL04,JaveyNat03,AvourisMT06,AvourisNNT07} 
and the hope that these carry over to high frequencies.
Measurements of the high-frequency characteristics of carbon nanotube
field-effect transistors (NTFETs)
\cite{AppenzellerAPL04,LiNL04,RojasNL07,ChasteNL08} indicate little decrease
in performance up to gigahertz (GHz) frequencies.
In addition, the carrier dynamics in the terahertz (THz) regime was recently
probed using time-domain techniques, suggesting that the carrier dynamics is
determined by single-particle rather than plasmonic properties.\cite{ZhongNNT08}

Theoretically, the problem of time-dependent transport has been approached
using a variety of techniques such as scattering matrix theory,
\cite{BuettikerPRL93,BuettikerJPC93,PretrePBR96,PedersenPRB98,WangAPL07}
Floquet methods,\cite{SambePRA73,BrandesPRB97,MartinezJPA03,CamaletPRL03,
IndlekoferPRB08,WuJPC08,HoPRB09}
Boltzmann transport theory,\cite{AkturkPRL07,AkturkJAP07,PaydavosiTNT09,PaydavosiTNT09-2}
and nonequilibrium Green Functions
(NEGF).\cite{JauhoPRB93,JauhoPRB94,YouPRB00,FranssonIJCP03,StefanucciPRB04,KurthPRB05,
ZhouJPD05,ZhuPRB05,StefanucciJPCS06,HouPhysE06,MoldoveanuPRB07,MoldoveanuPRB07-2, SouzaPRB07,
StefanucciPRB08,MaciejkoPRB06,MyoehaenenEPL08,MyoehaenenPRB09,StanJCP09, PanJPC09,DattaPRB92,
AnantramPRB95,GuoPRL99,RolandPRL00,WangPRB03,WeiPRB09,ZhengPRB00, WuPRB05,YuJCP07,WangPRB09}
%	TD-NEGF --> Time-Domain: Quantum Dots etc.
Even though the NEGF technique has become to some extent the standard in
modeling electronic quantum transport, its application to time-dependent
problems has been mainly focused on simplified few-level models,
\cite{DattaPRB92,AnantramPRB95,GuoPRL99,RolandPRL00,ZhengPRB00,WuPRB05,YuJCP07,WangPRB09}
or to few-atom one-dimensional wires or molecules.\cite{ZhengPRB00,WuPRB05,YuJCP07,WangPRB09}
While such model systems are invaluable to gain insight into the basic dynamic
processes of simple quantum systems, it can be difficult to relate them to more
realistic devices for three main reasons.
First, the self-consistency between the charge and the potential is needed to
properly determine the operation of the device in the presence of dynamic
potentials. Second, it is necessary to capture the rather complex dielectric
environment of real devices. Third, most approaches have focused on applying 
time-dependent signals at the source-drain, \textit{i.e.} transport terminals,
\cite{FranssonIJCP03,MoldoveanuPRB07,MoldoveanuPRB07-2,SouzaPRB07}
rather than at the gate, a \textit{non-transport} terminal.
Physically, non-transport terminals do not supply the device region with
charge, unlike source-drain contacts, but are coupled to the device channel
only through the (self-consistent) dynamic potential, which plays a similar
role as the pumping potential in the field of parametric
pumping.\cite{StefanucciPRB08,WangPRB03}

In this work, we make a first step towards solving this problem and develop
a linear response theory for ac quantum transport employing nonequilibrium
Green functions solved self-consistently with Poisson's equation, when a
time-dependent signal is applied at the gate terminal. We apply the approach
to a NTFET and determine its high-frequency response, showing that the
approach cannot only describe time-dependent, single-particle quantum
transport effects, but also is able to capture the plasmonic excitations
of the device.

The program of the paper is as follows: in Sec. II, we detail the
formal theory for ac quantum transport and derive an effective Dyson
equation describing the dynamics of the system for a time-harmonic
signal at a non-transport terminal. Special attention is given to the
practical calculation of the frequency-dependent charge density for which
we develop a computationally efficient scheme, a prerequisite for calculating
the self-consistent ac response of larger systems, as we have demonstrated
previously.\cite{KienlePRL09}
In Sec. III, general expressions for the ac particle current and associated
conductance are derived along with a brief outline of the current partitioning scheme,
\cite{GuoPRL99}
and how it applies to a multi-terminal device with non-transport terminals.
In Sec. IV, we apply the theory to a NTFET. There, we discuss details of the
significance of the operation mode of the device, and the self-consistent feedback
between charge and potential for collective excitations.
Our conclusions are presented in Sec. V.

\section{GENERAL APPROACH}
In this section we describe the development of the ac approach, which
consists of three steps: 1) definition of the model Hamiltonian of the total
system, 2) formulating the quantum dynamics and non-equilibrium statistics
in terms of Green functions in the energy domain, and 3) self-consistent
calculation of the ac charge and potential.

\subsection{Model Hamiltonian}
We begin by specifying the Hamiltonian operator of the system. As usual, the
total system is divided into three isolated regions following the partitioning
scheme of Caroli and co-workers\cite{CaroliJPC71-1,CaroliJPC71-2}
The Hamiltonian of the entire infinite system is written as
\begin{eqnarray}\label{EqHamTot}
H=H_{d}+H_{c}+H_{t}~,
\end{eqnarray}
where $H_{d}$ is the Hamiltonian for the device region, $H_{c}$ refers to the
two semi-infinite leads, and $H_{t}$ couples the device region to the leads.
In a site representation the device Hamiltonian is given by 
\begin{eqnarray}
H_{d} &=& H_{d}^{0} + H_{d}^{DC} + H_{d}^{AC}~,
\end{eqnarray}
where
\begin{eqnarray}
H_{d}^{0} &=& \sum_{n} \epsilon_{n}^{0} \hat{c}_{n}^{\dag}\hat{c}_{n}
+\frac{1}{2} \sum_{n,m} t_{n,m} \hat{c}_{n}^{\dag} \hat{c}_{m}+\mbox{h.c.}~,
\label{Eq:HamDev0}
\end{eqnarray}
and 
\begin{eqnarray}
H_{d}^{DC} &=& \sum_{n} U_{n}^{DC} \hat{c}_{n}^{\dag}\hat{c}_{n}~,~
H_{d}^{AC} = \sum_{n} U_{n}(t) \hat{c}_{n}^{\dag}\hat{c}_{n}~, \label{Eq:HamDevU}
\end{eqnarray}
where $\hat{c}_{n}^{\dag}$ and $\hat{c}_{n}$ refer to fermionic creation and
annihilation operators at site $n$.
$H_{d}^{0}$ defines the equilibrium electronic structure of the isolated system.
The electron-electron interaction is approximated on the Hartree level and has
two components $H_{d}^{DC}$ and $H_{d}^{AC}$, cf. Eq~(\ref{Eq:HamDevU}).
The term $U_{n}^{DC}$ represents a spatially-varying, but time-\textit{independent}
electrostatic potential, such as the one present when calculating the DC properties
and leads to a renormalization of the onsite energies $\epsilon_{n}^{0}$.

The new physics studied here originates from the presence of an \textit{a priori unknown}
time- and space-\textit{dependent} potential $U_{n}(t)$ induced by externally applied
time-dependent fields. As further discussed below, both $U_{n}^{DC}$ and $U_{n}(t)$ must
be determined separately by solving Poisson's equation in a self-consistent manner.
In general, the approach allows to investigate the dynamic response beyond the Hartree
approximation of the Coulomb interaction by including exchange and correlation
functionals\cite{WangPRB09} calculated self-consistently.

The Hamiltonian for the two contacts to the left and right ($\alpha = s,d$)
of the device reads 
\begin{equation}\label{EqHamCon}
H_{c} = \sum_{k,\alpha} \epsilon_{k\alpha}^{0} 
\hat{c}_{k\alpha}^{\dag}\hat{c}_{k\alpha}~,
\end{equation}
where $\hat{c}_{k\alpha}^{\dag}$ and $\hat{c}_{k\alpha}$ are fermionic
creation and anihilation operators for a particle in terminal $\alpha$
in state $k$.
We note that equations (\ref{Eq:HamDev0}), (\ref{Eq:HamDevU}), and (\ref{EqHamCon})
differ from those considered previously where a time-dependent source-drain bias is
considered, in which case the onsite energy $\epsilon_{k\alpha}^{0}$ of the
contacts become time-dependent rather than the ones of the device.

Finally, the Hamiltonian 
\begin{equation}\label{EqHamTun}
H_{t} = \sum_{k\alpha,n}
T_{n,k\alpha}        \hat{c}_{n}^{\dag}\hat{c}_{k\alpha} + 
T_{n,k\alpha}^{\ast} \hat{c}_{k\alpha}^{\dag}\hat{c}_{n}~,
\end{equation}
couples the device sub-space with the semi-infinite source and drain reservoirs,
and allows for a physical exchange of particles through the device-contact
interface. Therefore, the tunneling Hamiltonian Eq.~(\ref{EqHamTun}) describes
only the coupling between the device and \textit{transport} terminals, but not
to \textit{non-transport} terminals.

\subsection{Quantum dynamics and nonequilibrium statistics}
The next step is to describe the carrier dynamics within the device
scattering region using Green functions. The Green functions are in
general functions of both space and time, \textit{e.g.}
$G(\mathbf{r}t;\mathbf{r}^{\prime} t^{\prime })$.
However, to simplify the equations for compactness we adopt a short-hand
notation
$G(t,t^{\prime}) \equiv G(\mathbf{r} t;\mathbf{r}^{\prime} t^{\prime })$.
In addition, whenever regular functions appear with Green functions in the
same equation, we also omit the spatial dependence on the regular functions.

We start with the time-dependent Dyson equation\cite{Haug1998} 
\begin{eqnarray}\label{EqDys1}
G^{\gamma }(t,t^{\prime }) &=& g_{0}^{\gamma }(t,t^{\prime }) \\
&+&\int dt_{1}dt_{2}~g_{0}^{\gamma }(t,t_{1})
\Sigma^{\gamma}(t_{1},t_{2}) G^{\gamma }(t_{2},t^{\prime})~, \nonumber
\end{eqnarray}
where $g_{0}^{\gamma}(t,t^{\prime }) = g_{0}^{\gamma}(t-t^{\prime })$
refers to the \textbf{r}etarded/\textbf{a}dvanced
($\gamma =\mbox{\textbf{r},\textbf{a}}$) Green function of the \textit{isolated}
system. The self-energy $\Sigma ^{\gamma}(t,t^{\prime})$ accounts for all
interactions of the isolated system with its environment.
In our case, the self-energy $\Sigma ^{\gamma }(t,t^{\prime })$
can be divided into three contributions 
\begin{equation}\label{EqSig1}
\Sigma ^{\gamma }(t,t^{\prime }) = \sum_{\alpha=s,d}
\Sigma_{\alpha }^{\gamma}(t-t^{\prime })
+ U^{DC} \delta (t-t^{\prime })
+ U(t) \delta (t-t^{\prime })~.
\end{equation}
The first term $\Sigma_{c} \equiv \sum_{\alpha=s,d} \Sigma_{\alpha}$ is the
contact self-energy and corresponds to the quantum-transport open-boundary
conditions connecting the device region with the semi-infinite source and
drain contacts. The second term is a scalar potential and represents the
internal response of the device to externally applied time-\textit{independent}
fields. The third term is the prominent feature in the ac theory presented here,
and describes the \textit{dynamic} response of the device due to external
time-\textit{dependent} fields. Contrary to most studies where the ac signal
is applied at the source-drain
terminals,\cite{GuoPRL99,ZhouJPD05,WangPRB09,WeiPRB09} in our case the
time-dependent signal is applied at the gate terminal. This implies that
the induced potential $U(t)$ distorts only the device scattering region,
while the contacts remain in steady-state.

We now switch from the time-domain into energy-representation through a
double-time Fourier-transform defined as\cite{GuoPRL99}
\begin{equation}\label{EqFourier}
F(E,E^{\prime }) = \int dtdt^{\prime }
e^{iEt/\hbar } e^{-iE^{\prime }t^{\prime}/\hbar } F(t,t^{\prime })
\end{equation}
and
\begin{equation}
F(t,t^{\prime}) = \int \frac{dE}{2\pi}\frac{dE^{\prime}}{2\pi}
e^{-iEt/\hbar} e^{iE^{\prime}t^{\prime}/\hbar} F(E,E^{\prime})~,
\end{equation}
so that the self-energy, cf. Eq.~(\ref{EqSig1}) is given by 
\begin{equation}\label{EqSig3}
\Sigma^{\gamma}(E,E^{\prime}) =
2\pi \delta (E-E^{\prime})
\left[ \Sigma_{c}^{\gamma}(E) + U^{DC} \right] + U(E-E^{\prime}).
\end{equation}
It is worthwhile mentioning that in energy domain the contact self-energies
are local in energy, reflecting that under \textit{steady-state} conditions
there is no mixing between states with different energy within the reservoirs.
On the other hand, the original time-local potential $U$ becomes now in energy
domain \textit{non-local}, implying that a time-dependent potential mediates
transitions between states at different energies within the device scattering
region.

Fourier transforming Eq.~(\ref{EqDys1}) and using (\ref{EqSig3}), one
derives an effective Dyson equation for the device
\begin{eqnarray}
G^{\gamma}(E,E^{\prime}) &=&
2\pi \delta (E-E^{\prime}) G_{0}^{\gamma}(E) \label{EqDys2:1} \\
&+& \int \frac{d\bar{E}}{2\pi}~G_{0}^{\gamma}(E)
U(E-\bar{E}) G^{\gamma}(\bar{E},E^{\prime})~, \nonumber
\end{eqnarray}
where
\begin{eqnarray}
G_{0}^{\gamma}(E) &=& 
\left[ g_{0}^{\gamma}(E)^{-1} - U^{DC} - 
\Sigma _{c}^{\gamma}(E) \right]^{-1} ~,\label{EqDys2:2}
\end{eqnarray}
and
\begin{eqnarray}
g_{0}^{\gamma}(E) &=& \left[ \left(E \pm i\eta\right) I
- H_{d}^{0} \right]^{-1}~\label{EqDys2:3},
\end{eqnarray}
with an infinitesimal $\eta>0$.
What we have gained in re-formulating Dyson's equation is to partition the
full dynamic response of the system described through the two-energy Green
function $G(E,E^{\prime})$ into its dc and ac components given by the first
and second term in equation (\ref{EqDys2:1}), respectively.
Importantly, the DC component determined by the newly defined Green function
$G_{0}^{\gamma}$, cf. Eq.~(\ref{EqDys2:2}), refers no longer to the 
response of the isolated system $g_{0}^{\gamma}$, cf. Eq.~(\ref{EqDys2:3}),
but rather describes the system's response in contact with the leads and subject 
to a dc electrostatic potential. Hence, $G_{0}^{\gamma}$ defines the operation
point of the \textit{open} system under DC steady-state.
The ac component, \textit{i.e.} the second term in Eq.~(\ref{EqDys2:1}) contains
this term as well and determines the \textit{distortion} of the system away from
the operation point $G_{0}^{\gamma}$, and is driven by the time-dependent potential
$U(t)$ leading to a coupling of states at different energies.

We still need to know how the total nonequilibrium particle distribution
$G^{<}$ deviates from its (reference) distribution at dc in the presence of
the ac potential $U$. This is accomplished by mapping Dyson's equation for
$G^{<}$, symbolically written as
$G^{<}=\left[ G_{0} + G_{0} U G \right] ^{<}$, 
onto the real-time axis utilizing the Langreth rules\cite{LangrethPRB72,Haug1998}
which gives: $G^{<}=G_{0}^{<}+G_{0}^{<}UG^{a}+G_{0}^{r}UG^{<}$.
This integral equation can be solved exactly making use of Eqs.~(\ref{EqDys2:1})
and (\ref{EqDys2:2}). Details of the derivation are found in appendix A.
After Fourier transform the particle distribution is given by 
\begin{eqnarray}\label{EqGL}
G^{<}(E,E^{\prime }) &=& 2\pi G_{0}^{<}(E)\delta (E-E^{\prime})  \\
&+& \int \frac{d\bar{E}}{2\pi }
\left[ 
G_{0}^{<}(E) U(E-\bar{E}) G^{a}(\bar{E},E^{\prime }) \right. \nonumber \\
&+& \left. ~~~~~~~~~~G^{r}(E,\bar{E}) U(\bar{E}-E^{\prime}) G_{0}^{<}(E^{\prime })
\right] \nonumber \\
&+& \int \frac{dE_{1}}{2\pi }\frac{dE_{2}}{2\pi }\frac{dE_{3}}{2\pi }\times 
\nonumber \\
&&~~~~G^{r}(E,E_{1}) U(E_{1}-E_{2}) G_{0}^{<}(E_{2}) \times  \nonumber \\
&&~~~~U(E_{2}-E_{3}) G^{a}(E_{3},E^{\prime })~,\nonumber
\end{eqnarray}
where
$G_{0}^{<}(E) = G_{0}^{r}(E) \Sigma _{c}^{<}(E) G_{0}^{a}(E)$
corresponds to the nonequilibrium spectral particle density at dc.
The function $\Sigma_{c}^{<}(E)=\sum_{\alpha=s,d} if_{\alpha}(E)\Gamma_{\alpha}(E)$
where
$\Gamma_{\alpha}(E)=i\left( \Sigma _{\alpha}^{r}-\Sigma_{\alpha}^{a}\right)$
is the broadening function, and
$f_{\alpha}(E) = 1 / \left[ 1+e^{(E-\mu_{\alpha})/k_{B} T}\right]$
is the Fermi function at temperature $T$ with $\mu_{\alpha}$ being the
chemical potential of terminal $\alpha$.

While the set of equations (\ref{EqDys2:1})-(\ref{EqGL}) developed so far
describe entirely the quantum transport and nonequilibrium statistics, they
do not allow to determine the dynamic potential $U$. This must be obtained by
solving Poisson's equation 
\begin{equation}\label{EqPoi1}
\nabla \left[ \epsilon (\mathbf{r})\nabla U(\mathbf{r},E-E^{\prime })
\right] =-\rho (\mathbf{r},E-E^{\prime })~,  
\end{equation}
with the frequency-dependent charge density 
\begin{equation}\label{EqRho1}
\rho (\omega )=ie\int \frac{dE}{2\pi }~G^{<}(E^{+},E)~.  
\end{equation}
The calculation of the ac charge density using Eq.~(\ref{EqGL}) requires
$G^{<}$ to be evaluated at two energies $(E^{+},E)\equiv (E+\hbar \omega ,E)$,
in contrast to the dc case where only one energy is needed.

Equations~(\ref{EqPoi1}) and (\ref{EqRho1}) implement the \textit{self-consistent}
coupling between electrostatics and transport, which represents the key component
in our ac approach. Note that at the frequencies considered here the electromagnetic
fields respond instantaneously, so that the full time dependence in Maxwell's equations
can be neglected. Poisson's equation is supplemented by boundary conditions appropriate
for the problem at hand, and $\epsilon (\mathbf{r})$ is a space-dependent dielectric
constant that can account for more complex inhomogeneous dielectric environments quite
common in devices.

The set of Eqs.~(\ref{EqDys2:1})-(\ref{EqRho1}) describe the nonequilibrium quantum
dynamics and its coupling to Poisson's equation for an arbitrary time-dependent potential
$U$, and can thus describe situations beyond linear response, in general.
However, the numerical implementation of the full non-linear theory requires the calculation
of a triple energy integral in Eq.~(\ref{EqGL}), which is prohibitive at this time given the
need for self-consistency to capture the plasmonic response of real devices as discussed in
Sec. IV.

\subsection{Linearized equations}
To proceed further, we now apply a time-harmonic signal at the gate terminal 
${\tilde{v}}_{g}(t)=v_{0}\cos (\omega t)$ of small amplitude $v_{0}$ and
frequency $\omega $, and seek the potential response in the form
$U(\mathbf{r},t)=V(\mathbf{r},\omega )\cos (\omega t)$, which reads in energy
domain 
\begin{equation}\label{EqUE}
U(E) = \frac{1}{2}V(\mathbf{r},\omega) 
\left[ \delta(E+\hbar\omega) + \delta(E-\hbar\omega)\right]~.  
\end{equation}
Keeping only terms to linear order in $V$, the ac transport-Poisson equations
take the form 
\begin{eqnarray}
G^{\gamma }(E^{+},E) &=& 2\pi G_{0}^{\gamma }(E)\delta (\hbar \omega )
+\frac{1}{2} G_{0}^{\gamma }(E^{+}) V(\omega ) G_{0}^{\gamma }(E), \nonumber \\
\label{EqAC:GR} \\
G^{<}(E^{+},E) &=& 2\pi G_{0}^{<}(E) \delta(\hbar\omega)
+   \frac{1}{2} G_{0}^{<}(E^{+}) V(\omega) G_{0}^{a}(E)  \nonumber \\
&+& \frac{1}{2} G_{0}^{r}(E^{+}) V(\omega) G_{0}^{<}(E)~, \label{EqAC:GL} \\
\rho (\omega ) &=&ie\int \frac{dE}{2\pi }~G^{<}(E^{+},E), \label{EqAC:Rho} \\
-\rho (\mathbf{r},\omega ) &=&\nabla\left[ \epsilon (\mathbf{r})
\nabla V(\mathbf{r},\omega ) \right]~.  \label{EqAC:Poi}
\end{eqnarray}

\subsection{Numerical calculation of $\rho(\mathbf{r},\omega)$}
An integral part in the self-consistent transport calculations is the
determination of the charge density. In practice, one has to evaluate
the integral in Eq.~(\ref{EqAC:Rho}) which is often performed by direct
integration along the real energy axis. In many cases, this is a sufficient
approach because the spectral density of states has a finite bandwidth, thus
narrowing the integration window. However, such conditions are rarely realized
in more realistic device models. For instance, even in a simple tight-binding
representation of a NTFET (see Sec. IV) the bandwidth of the valence and
conduction band is about $10$ eV, in which case the calculation of the charge
density through a real-axis integration can become prohibitive for self-consistent
calculations even at dc. This becomes an even more severe bottleneck in the case of
ac simulations, where now the charge has to be determined at every frequency $\omega$.

In the following, we describe a computational efficient approach, which permits
the calculation of the \textit{frequency-dependent} charge density $\rho(\omega)$
by exploiting contour integration in the complex energy
plane.\cite{ZellerSSC82,BrandbygePRB02}
The basic idea is similar to the dc case, \textit{i.e.} to separate in
Eq.~(\ref{EqAC:Rho}) the zero-bias contribution to $\rho(\omega)$ from the
non-zero-bias component. If we further assume that the lowest chemical potential
is at the drain terminal, \textit{i.e.} $\mu_{d}^{(+)} < \mu_{s}^{(+)}$ the
frequency-dependent particle distribution at \textit{zero-bias} (ZB) reads
\begin{eqnarray}\label{EqRhoW:1}
G_{ZB}^{<}(E^{+},E) &=& 
  \frac{i}{2} A_{0}^{+} V(\omega) G_{0}^{a} f_{d}^{+} 
+ \frac{i}{2} G_{0}^{r,+} V(\omega) A_{0} f_{d}~,\nonumber \\
\end{eqnarray}
where a $+$ superscript indicates a function evaluated at $E+\hbar \omega$,
and the absence of such a superscript indicates a function evaluated at $E$.
The steady-state spectral density is given by
\begin{equation}\label{Eq:A0}
A_{0}^{(+)} = i\left[ G_{0}^{r,(+)} - G_{0}^{a,(+)}\right]~.
\end{equation}
$G_{ZB}$ contains Fermi functions evaluated a two different energies
$E$ and $E^{+}$, reflecting the nonequilibrium nature of the ac
charge density for finite frequencies even at zero-bias, which means
that an externally applied ac signal acts as if a frequency-dependent
bias were applied.

Taking advantage of this ac-signal-bias analogy and noting that $f_{d}^{+}<f_{d}$,
one can split again the zero-bias particle density into
its equilibrium and nonequilibrium components,
\textit{i.e.} $G_{ZB}^{<} = G_{ZB}^{<,eq} + G_{ZB}^{<,neq}$, which after
re-arrangement take the form 
\begin{equation}\label{Eq:GZBeq}
G_{ZB}^{<,eq}=-\frac{i}{2}\left( f_{d}+f_{d}^{+}\right) ~\mbox{Im}\left\{
G_{0}^{r,+}V(\omega )G_{0}^{r}\right\} ~,
\end{equation}
and
\begin{eqnarray}\label{Eq:GZBneq}
G_{ZB}^{<,neq} &=&
-\frac{1}{2}\left( f_{d}-f_{d}^{+}\right)
\left[ \mbox{Re}\left\{ G_{0}^{r,+} V(\omega ) G_{0}^{r}\right\} \right.  \\
&&~~~~~~~~~~~~~~~~~~~~~~~\left. -G_{0}^{r,+} V(\omega ) G_{0}^{a}\right] ~. 
\nonumber
\end{eqnarray}
The equilibrium part $G_{ZB}^{<,eq}$ is \textit{analytic} in the upper
complex plane, since it consists of the product of two retarded Green
functions $G_{0}^{r}$ and $G_{0}^{r,+}$ each of which has poles only in
the lower complex plane.\cite{Haug1998}
Therefore, the equilibrium zero-bias ac particle density, which involves
all states below the frequency-dependent chemical potential $\mu_{d}^{+}$,
can be efficiently calculated through integration over a complex energy
contour.\cite{ZellerSSC82,BrandbygePRB02}
Conversely, the nonequilibrium component at zero-bias $G_{ZB}^{<,neq}$
is \textit{non-analytic}, because both the retarded and the advanced Green
functions are needed with their corresponding poles located in the lower
and upper complex plane, respectively. However, this does not pose a serious
problem in practice as the integration range is limited to a finite energy
window given by $\hbar\omega$, \textit{i.e.} the difference between the chemical
potentials $\mu_{d} - \mu_{d}^{+}$.\cite{Note3}

\section{ac RESPONSE FUNCTIONS: CURRENT AND CONDUCTANCE}
The set of equations (\ref{EqAC:GR})-(\ref{EqAC:Poi}) developed in the
previous section allow the determination of the frequency-dependent Green
functions, which can now be used to obtain ac response functions. One basic
response function to characterize transport is the dynamic conductance
$g_{\alpha\beta}$, which relates the total ac current $I_{\alpha}$ with the
voltage applied at terminal $\beta$.
Under time-dependent conditions this conductance is not entirely determined by
the particle current, but has in general contributions from the displacement
current as well. In the following subsections, we derive an expression for the
particle conductance, and summarize how displacement currents can be included in
the total conductance.

\subsection{Particle current $I_{\alpha}^{p}(\omega)$}
The first contribution to the total current consists of the flow of charged
particles through the terminal $\alpha$, and is hence determined by the dynamic
change of the particle density at this terminal
\begin{eqnarray}\label{EqIpT1}
I_{\alpha}^{p}(t) = -e\frac{d}{dt} \langle \hat{N}_{\alpha}(t)\rangle
= -e\frac{d}{dt}\sum_{k}
\langle\hat{c}_{k\alpha}^{\dag}(t)\hat{c}_{k\alpha}(t)\rangle~.
\end{eqnarray}
Making use of the fermionic anti-commutator
relations,\cite{Haug1998} and the Heisenberg equation of motion for operators
$\dot{\mathcal{O}} = \frac{i}{\hbar}\left[ H,\mathcal{O} \right]$ with $H$ the
total system Hamiltonian, cf. Eq.~(\ref{EqHamTot}), one derives the well-known
matrix equation for the equal-time particle
current\cite{JauhoPRB94,AnantramPRB95,WeiPRB09} 
\begin{eqnarray}\label{EqIpT2}
I_{\alpha}^{p}(t) &=&\frac{e}{\hbar}Tr\int dt^{\prime}
\left[
G^{r}(t,t^{\prime}) \Sigma_{\alpha}^{<}(t^{\prime},t)
- \Sigma_{\alpha}^{<}(t,t^{\prime}) G^{a}(t^{\prime },t) \right. \nonumber \\
&+& \left. G^{<}(t,t^{\prime}) \Sigma_{\alpha}^{a}(t^{\prime},t)
-\Sigma_{\alpha}^{r}(t,t^{\prime}) G^{<}(t^{\prime},t)\right] ~.
\end{eqnarray}
Its corresponding energy representation
reads\cite{JauhoPRB94,AnantramPRB95,GuoPRL99} 
\begin{eqnarray}\label{EqIpE}
I_{\alpha}^{p}(\omega) &=& \frac{e}{\hbar}
Tr\int\frac{dE}{2\pi} \frac{dE^{\prime}}{2\pi} 
\left[
G^{<}(E^{+},E^{\prime}) \Sigma_{\alpha}^{a}(E^{\prime },E)\right.  \nonumber \\
&-& \left. \Sigma_{\alpha}^{r}(E^{+},E^{\prime }) G^{<}(E^{\prime},E)
+ G^{r}(E^{+},E^{\prime}) \Sigma_{\alpha }^{<}(E^{\prime },E)\right. \nonumber \\
&-& \left. \Sigma_{\alpha}^{<}(E^{+},E^{\prime}) G^{a}(E^{\prime },E)\right]~.
\end{eqnarray}
Equation~(\ref{EqIpE}) simplifies further if we exploit the steady-state
property of the contact self-energies from Eq.~(\ref{EqSig3}) in which case
one obtains the frequency-dependent particle current
\begin{eqnarray}\label{EqIpE2}
I_{\alpha}^{p}(\omega ) &=& \frac{e}{h}Tr\int dE
\left[
G^{<}(E^{+},E) \Sigma_{\alpha}^{a}(E)\right.  \nonumber \\
&-& \left. \Sigma_{\alpha}^{r}(E^{+}) G^{<}(E^{+},E) 
+ G^{r}(E^{+},E) \Sigma_{\alpha}^{<}(E) \right.  \nonumber \\
&-& \left. \Sigma_{\alpha}^{<}(E^{+}) G^{a}(E^{+},E)\right] ~.
\end{eqnarray}
We note that the expression for $I_{\alpha}^{p}(\omega )$ differs from
those derived in Refs.\cite{AnantramPRB95,GuoPRL99,WeiPRB09}, since those
applied a time-dependent voltage at the source-drain, instead of the gate
excitation considered here.

One can now derive the dynamic particle conductance by expanding
$\Sigma_{\alpha}^{<}$ and $G^{<}$ appearing in Eq.~(\ref{EqIpE2}) to linear
order in the terminal voltage $v_{\beta}$, and utilizing Eq.~(\ref{EqAC:GR})
and (\ref{EqAC:GL}) to substitute for $G^{\gamma}(E^{+},E)$ and $G^{<}(E^{+},E)$.
These linearized expressions are summarized in Appendix B. Inserting all
relevant terms in Eq.~(\ref{EqIpE2}) and keeping components linear in $V(\omega)$,
we derive for the frequency-dependent particle current 
\begin{widetext}
\begin{eqnarray}\label{EqIpE3}
I_{\alpha}^{p} (\omega) &=& 
\frac{1}{2}\frac{e^2}{h} \sum_{\beta} Tr \int dE
\left[
\left\{ G_{0}^{r,+} V(\omega) G_{0}^{r} \widetilde{\Sigma}_{\beta}^{<} 
- \widetilde{\Sigma}_{\beta}^{<,+} G_{0}^{a,+} V(\omega) G_{0}^{a} \right\} 
\delta_{\alpha\beta} \right. \\
&+& \left. \widetilde{G}_{0,\beta}^{<,+} V(\omega) G_{0}^{a} \Sigma_{\alpha}^{a} 
+ G_{0}^{r,+} V(\omega) \widetilde{G}_{0,\beta}^{<} \Sigma_{\alpha}^{a}
- \Sigma_{\alpha}^{r,+} \widetilde{G}_{0,\beta}^{<,+} V(\omega) G_{0}^{a} 
- \Sigma_{\alpha}^{r,+} G_{0}^{r,+} V(\omega) \widetilde{G}_{0,\beta}^{<}
\right]~v_{\beta} ~~. \nonumber
\end{eqnarray}
\end{widetext}
By definition, the (tensor) prefactor that relates the terminal current
$I_{\alpha}$ with the applied bias $v_{\beta}$ is the \textit{ac linear response}
particle conductance, and can be read directly from Eq.~(\ref{EqIpE3}): 
\begin{widetext}
\begin{eqnarray}\label{Eq:GacPart}
g_{\alpha\beta}^{p}(\omega) &=&
\frac{1}{2}\frac{e^2}{h} Tr \int dE
\left[
\left\{ G_{0}^{r,+} V(\omega) G_{0}^{r} \widetilde{\Sigma}_{\beta}^{<} 
- \widetilde{\Sigma}_{\beta}^{<,+} G_{0}^{a,+} V(\omega) G_{0}^{a} \right\} 
\delta_{\alpha\beta} \right. \\
&+& \left. \widetilde{G}_{0,\beta}^{<,+} V(\omega) G_{0}^{a} \Sigma_{\alpha}^{a} 
+ G_{0}^{r,+} V(\omega) \widetilde{G}_{0,\beta}^{<} \Sigma_{\alpha}^{a}
- \Sigma_{\alpha}^{r,+} \widetilde{G}_{0,\beta}^{<,+} V(\omega) G_{0}^{a} 
- \Sigma_{\alpha}^{r,+} G_{0}^{r,+} V(\omega) \widetilde{G}_{0,\beta}^{<}
\right]~. \nonumber
\end{eqnarray}
\end{widetext}

\subsection{Displacement current $I^{d}(\omega)$}
Under time-dependent conditions the particle conductance does not in general obey
sum-rules, \textit{i.e.}
$\sum_{\alpha} g_{\alpha\beta} = 0$ and $\sum_{\beta} g_{\alpha\beta} = 0$,
reflecting current continuity and gauge-invariance, because the displacement
current present under ac conditions is often discarded.
The current partitioning scheme of Wang \textit{et al.}\cite{GuoPRL99} allows
to re-establish these sum-rules by taking displacement currents into account.

The basic idea of this scheme can be summarized as follows: starting from the 
charge continuity equation, $\partial_{t}\rho + \nabla \cdot \mathbf{j}^{p}=0$,
and integrating over the volume one obtains Kirchoff's current law,
\textit{i.e.} $I^{d}(t) + \sum_{\alpha} I_{\alpha}^{p} = 0$.
$I_{\alpha}^{p}$ refers to the particle current through terminal $\alpha$, and
can be associated with a particle conductance through
$I_{\alpha}^{p} = \sum_{\beta} g_{\alpha\beta}^{p} v_{\beta}$ with $v_{\beta}$
the voltage at terminal $\beta$.
The displacement current $I^{d}(t)=\partial_{t} Q(t)$ accounts for the dynamic
change of the total charge, and is non-zero under time-dependent conditions.

To obtain an expression for the total conductance defined by
$I_{\alpha} = \sum_{\beta} g_{\alpha\beta} v_{\beta}$ one needs to know how the
current $I_{\alpha}$ is split between the particle and the displacement current
at each terminal. While the particle component $I_{\alpha}^{p}$ is directly
accessible through transport, this is not immediately possible for $I^{d}$,
since only the \textit{total} rather than the \textit{terminal} displacement
current is known. This problem can be resolved by making two \textit{Ans\"{a}tze}
for the terminal and total displacement current,\cite{GuoPRL99}
\textit{i.e.}
$I_{\alpha} := I_{\alpha}^{p} + A_{\alpha} I^{d}$
and
$I^{d} := \sum_{\beta} g_{\beta}^{d} v_{\beta}$,
where $g_{\beta}^{d}$ defines the \textit{displacement} conductance, and
permits to specify a total conductance:
$g_{\alpha\beta} = g_{\alpha\beta}^{p} + A_{\alpha} g_{\beta}^{d}$.
The partitioning factor $A_{\alpha}$ can be determined by employing
the sum-rules $\sum_{\alpha} g_{\alpha\beta}=0$ and
$\sum_{\beta} g_{\alpha\beta}=0$,
so that the total conductance is given by\cite{GuoPRL99}
\begin{equation}\label{Eq:GacTot}
g_{\alpha\beta} = g_{\alpha\beta}^{p}
- \frac{\sum_{\gamma} g_{\alpha\gamma}^{p}}{\sum_{\gamma}g_{\gamma}^{d}} g_{\beta}^{d}~,  
\end{equation}
and constitutes a $(N \times N)$-matrix for a system with $N$-terminals, in general.

\section{APPLICATION: NANOTUBE FET}
\begin{figure}[tbph]
\centering
\includegraphics[width=8cm]{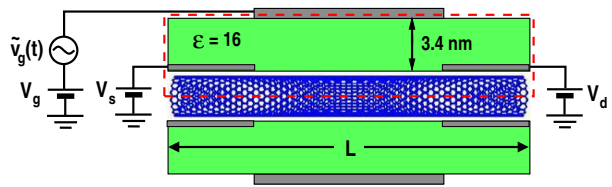}
\vspace{-0.2cm}
\caption{(Color online). Cross section of the NTFET device in cylindrical
geometry and embedded in a dielectric medium. The dashed rectangle specifies
the computational domain.}
\label{SketchDevice}
\end{figure}

In this section, we apply the approach developed in the previous sections to
a ballistic nanotube transistor shown in Fig.~\ref{SketchDevice} with a channel
length of $L=20$ nm. In general, the high-frequency properties of this three-terminal
device can be determined by any component of the
$(3 \times 3)$-conductance matrix, cf. Eq.~(\ref{Eq:GacTot}); here we chose the
source-drain conductance $g_{sd}(\omega)$ calculated at zero-bias and $T=300$ K.
Since the ac signal is applied only at the gate terminal, which couples capacitively
to the nanotube channel, one can assume that the dominant
contribution to the displacement current is carried by the gate,
\textit{i.e.} $I^{d}\approx I_{g}$.
In this case, the partitioning factors for the three terminals simplify,
\textit{i.e.} $(A_{s},A_{d},A_{g}) \approx (0,0,1)$, so that the source-drain ac
conductance is given by $g_{sd}(\omega) \approx g_{sd}^{p}(\omega)$.
Here, we focus on one particular channel length to discuss aspects of the methodology
that are essential for the proper description of the ac behavior. The properties of
such devices with different dimensions were presented by us in detail in Ref.~\cite{KienlePRL09}.

\subsection{Transistor response in the dc operation point}
We begin our analyis by specifying the setup of the device shown in Figure~\ref{SketchDevice} 
following the modeling approach of Ref.\cite{LeonardNT06}.
The channel, which consist of a semi-conducting tube with chirality $(m,n)=(17,0)$ and radius $0.66$ nm,
is placed in the center of a cylindrical hole with radius $0.96$ nm and surrounded by a dielectric with
a permittivity of $16$ ($\mbox{HfO}_2$). 

The equilibrium electronic structure of the nanotube is described within a $p_{z}$ tight-binding model
with diagonal matrix elements $\epsilon_{i,i}^{0} = 0$, and off-diagonal elements
$t_{2i,2i-1} = t_{2i-1,2i} = 2\gamma \cos \left( \frac{\pi J}{m}\right)$,
$t_{2i,2i+1} = t_{2i+1,2i} = \gamma$, where $m$ refers to the number of carbon atoms per ring. The periodic
boundary conditions along the tube circumference leads to a quantization of the wavefunction, so that the
NT electronic structure can be classified by an angular momentum $J=1,\ldots,m$ labeling the subbands.
We chose $\gamma = 2.5$ eV for the $\pi$ carbon-carbon bond energy, so that the bandgap between the highest
valence and lowest conduction band ($J=6$) is $E_{g}=0.55$ eV.\cite{LeonardNT06}

The contacts are semi-infinite extensions of the NT channel, and described through self-energies
$\Sigma^{r}_{\alpha}=\gamma^2 g_{\alpha}^{r}$ for each contact ($\alpha=s,d$) where $\gamma$ couples the first/last
ring of the NT channel to the surface of the contacts to the left and right.\cite{LeonardNT06}
The surface Green function $g_{\alpha}^{r}$ is calculated numerically at each energy using a matrix iterative
scheme.\cite{SanchoJPF85}
The matrix elements of the retarded Green function $G_{0}^{r}$ for the NT channel are obtained employing a
recursive algorithm.\cite{SvizhenkoJAP02}
The function of the two embedding metallic regions is to electrostatically dope the ends of the NT channel.
In all simulations, the equilibrium Fermi level of the semi-infinite NT source/drain contacts $E_{F}$
is set at $-1.0$ eV below the NT midgap energy before self-consistency, which gives after self-consistency
p-type Ohmic contacts.

Due to the cylindrical symmetry, the 3D Poisson's equation Eq.~(\ref{EqAC:Poi}) reduces to a two-dimensional
(2D) problem. In this case, Poisson's equation is discretized along the axial and radial axis within the 2D
simulation domain as marked by the rectangular box using finite-differences,\cite{LeonardNT06} and the resulting
linear matrix system is solved by successive overrelaxation.\cite{Press1992}
Along the domain boundary we impose homogeneous von Neumann boundary conditions for the electrostatic potential
($\nabla V=0$), and use Dirichlet boundary conditions ($V=const.$) at the perfect-metal source, drain, and gate
terminals. Poisson's equation requires a 3D charge density in real-space as input. However, an orthogonal
tight-binding representation of the NEGF transport equations calculates the total charge per NT ring.
A 3D charge density can be obtained by smearing of the total charge per ring along the axial and radial direction
of the 2D domain using Gaussian smearing functions.\cite{LeonardNT06}

The first step in determining the AC response of the transistor is to choose an operation point either in the off
or on state, which is controlled by an appropriate DC gate bias $V_{g}$.
\begin{figure}[tbph]
\centering
\includegraphics[width=8cm]{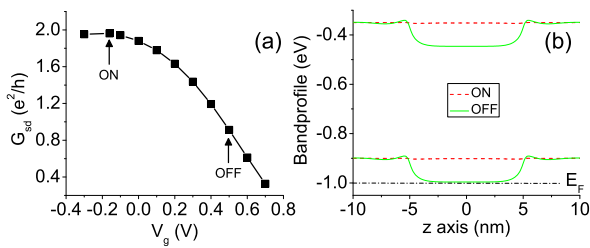}
\vspace{-0.4cm}
\caption{(Color online). (a) dc output characteristics for $L=20$ nm NTFET.
The arrows mark the operation point in the on and off state, while panel (b)
shows the respective self-consistent bandprofiles. The dashed-dotted line is
the Fermi level with $E_{F}=-1$ eV.}
\label{FigOutDC}
\end{figure}
Figure~\ref{FigOutDC} (a) shows the output characteristics for our NTFET specified
by the dc source-drain conductance $G_{sd}$ in the absence of a gate perturbation
($v_{0}=0$) with arrows marking the selected on and off states. At zero frequency,
the conductance $g_{sd}(\omega=0)$ examined in the forthcoming sections is related
to $G_{sd}$ by its slope taken at the operation point, \textit{i.e.}
$g_{sd}(\omega=0) \approx\frac{\partial G_{sd}}{\partial V_{g}} v_{0}$.
Figure~\ref{FigOutDC} (b) displays the respective dc bandprofiles with the
band being flat in the on state leading to a maximum conductance of $2e^{2}/h$
(per spin) whereas in the off state the hole current is reduced due to the
gate-controlled barrier in the channel.

\subsection{Transistor response in the off-state}
\begin{figure}[tbph]
\centering
\includegraphics[width=8cm]{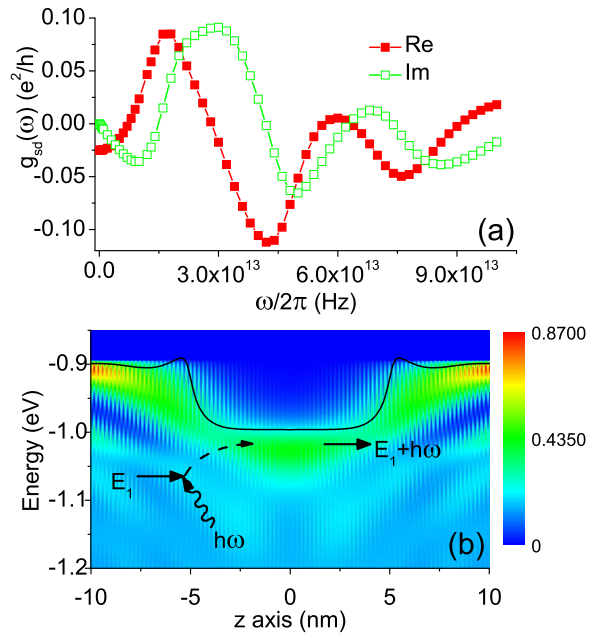}
\vspace{-0.4cm}
\caption{(Color online). ac response for a $L=20$ nm NTFET in the off state:
(a) real/imaginary part of the frequency-dependent conductance, and (b)
color plot of the NTFET 2D density-of-states illustrating the resonant
photoexcitation of carriers between spatially and energetically oscillating
states. The solid black line marks the valence band edge, and the Fermi
level is $E_{F}=-1$ eV.}
\label{FigOffState}
\end{figure}

We now superpose an ac signal of small amplitude $v_{0}=10$ meV and frequency
$\omega$ to the dc gate bias $V_{g}$. Figure~\ref{FigOffState} (a) shows the
dynamic conductance in the off state, with real and imaginary parts having
oscillatory character as a function of frequency $\omega$. One can understand
this behavior from the space- and energy-dependent 2D density-of-states (DOS)
shown in Fig~\ref{FigOffState} (b).
For a given position $z$ along the tube the DOS oscillates in energy due to the
quantum interference of states by the barriers. Photoexcitations of carriers between
states associated with maxima in the DOS lead to maxima in $g_{sd}(\omega)$, while
its minima are caused by transitions between maxima and minima.\cite{KienlePRL09}
An oscillatory behavior of the conductance is hence a signature of single-particle
excitations, and is preserved when the self-consistent feedback is disabled as will
be shown further below. We note that at low frequencies the real part of the conductance
is negative. This is because in the limit $\omega\rightarrow 0$ the ac signal perturbation
$\tilde{v}_{g}(t)=v_{0}\cos(\omega t)$ becomes effectively a positive DC bias
$\tilde{v}_{g}(t)=v_{0}>0$ superposed to $V_{g}$. According to the dc transfer
characteristics of Fig.~\ref{FigOutDC} (a) $\partial G_{sd}/\partial V_{g} < 0$,
so that an increase of $V_{g}$ by $v_{0}$ leads to a reduction in the conductance.

\subsection{Transistor response in the on-state}
\begin{figure}[tbph]
\centering
\includegraphics[width=8cm]{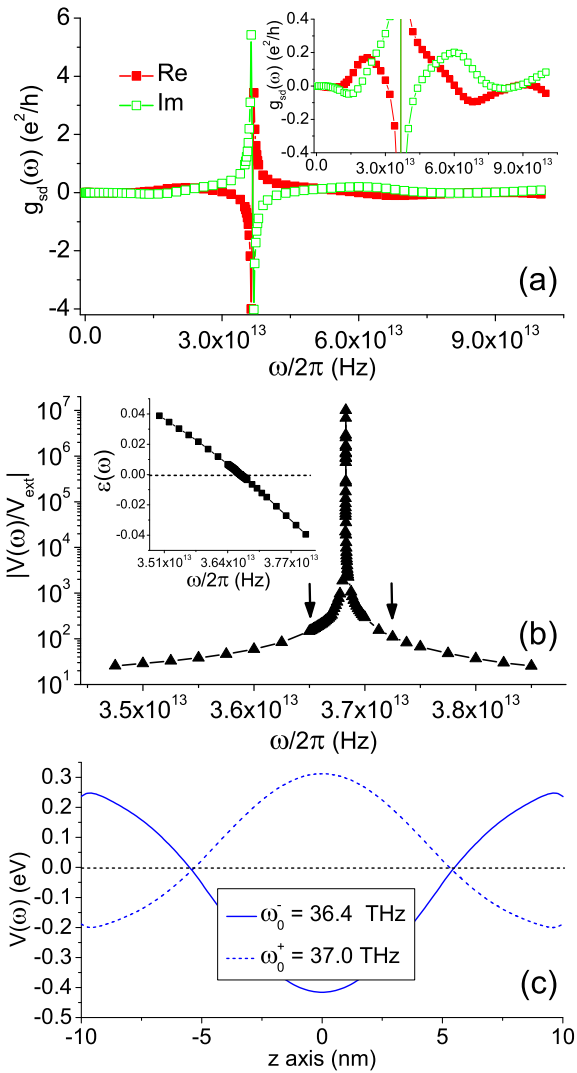}
\vspace{-0.4cm}
\caption{(Color online). ac response for a NTFET with $L=20$ nm in the on
state. Panel (a) shows the real/imaginary part of $g_{sd}(\omega)$ with
an enlarged view of the data shown in the inset. Panel (b): Amplitude of
the normalized potential $|V(\omega )/V_{ext}|$ with the dielectric function
$\epsilon(\omega)$ shown in the inset. The arrows mark the near-resonance
frequencies $\omega_{0}^{\pm }$. Panel (c) shows the large amplitude potential
profile $V(\omega)$ for $\omega_{0}^{\pm }$.}
\label{FigOnState}
\end{figure}

The dynamic response is quite different in the on state as shown in
Figure~\ref{FigOnState} (a). For small frequencies the dynamic conductance
is slightly negative for the same reason as in the off state, and exhibits
a pronounced divergence at a discrete frequency of about $\approx 36$ THz.
Away from this resonance the conductance is oscillatory similar to the off
state, as shown more clearly in the inset of Fig.~\ref{FigOnState} (a).

In order to identify the nature of this resonance, we determine the response
of the electrostatic potential $V(\omega)$ for fine-sampled frequencies
near the divergent behavior. In Figure~\ref{FigOnState} (b) we show the
ratio of $V(\omega)/V_{ext}$ with $V_{ext}$ the external perturbing
potential. Interestingly, upon approaching the resonance the amplitude of the
potential diverges, cf. Fig.~\ref{FigOnState} (b).
Alternatively, one can evaluate the frequency-dependent dielectric screening 
$\epsilon (\omega) = V_{ext}/V(\omega)$ shown in the inset of
Fig.~\ref{FigOnState} (b), which has a clear zero crossing at $\omega_{0}$,
while the potential undergoes a change in sign as displayed in
Fig.~\ref{FigOnState} (c). These observations verify that the divergent
behavior of the dynamic conductance observed in the on state is attributed
to the excitation of plasmons which co-exist with the single-particle
excitations.

\subsection{Importance of self-consistency}
In the previous section, we were able to identify the basic features in
the dynamic conductance such as the oscillatory and divergent characteristics
with the single-particle and collective behavior of the channel electrons,
and concluded that plasmons can only be excited if the device is operated
in the on state.
\begin{figure}[tbph]
\centering
\includegraphics[width=8cm]{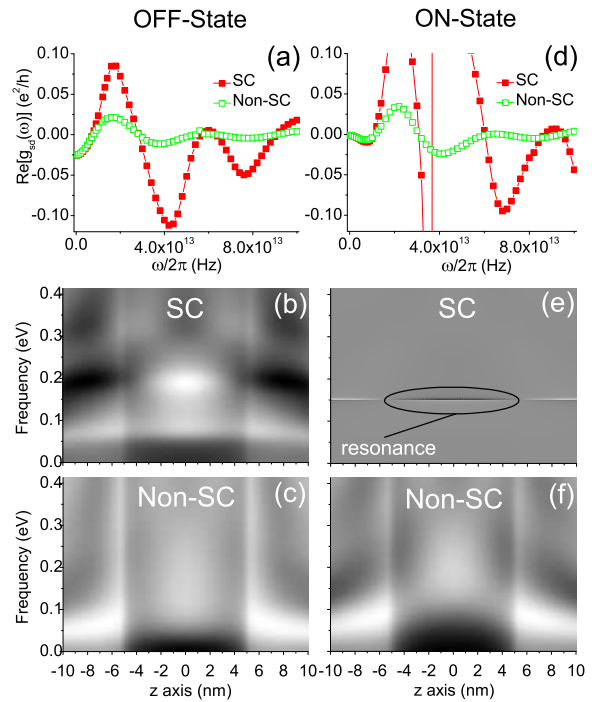}
\vspace{-0.4cm}
\caption{(Color online). Comparison of the self-consistent (SC) vs.
Non-self-consistent (Non-SC) response for a $L=20$ nm NTFET. Panels (a)-(c)
is the response in the off-state, with panel (a) the conductance and panels
(b) and (c) the charge density. Panels (d)-(e) show the behavior in the on-state.}
\label{FigSCF}
\end{figure}

Are there other, more fundamental prerequisites irrespective of the
operation point, which determine whether the system can be driven into a
collective state at all? The answer is yes, and is related to the
self-consistency between charge and potential. In Figure~\ref{FigSCF} we
compare the conductance $g_{sd}(\omega)$ and the frequency-dependent charge
density $\rho(z,\omega)$ calculated using the full self-consistent (SC)
and a non-self-consistent approach where in the latter case the dynamic
conductance is calculated in one step from the dc band profile.

In the off state, the most apparent difference between the SC vs non-SC
case is that, while the SC amplitude of the conductance is larger, the
smooth oscillatory behavior is preserved as shown in Fig.~\ref{FigSCF} (a).
Hence, the single-particle excitation spectrum is - at least qualitatively -
not affected by the charge-potential feedback. This is also apparent in the
ac charge density $\rho(z,\omega)$, cf. Figs.~\ref{FigSCF} (b) and (c),
which exhibits space- and frequency-dependent oscillations in both the
SC- and non-SC case.

In the on state, eliminating the feedback loop has a quite different impact
on the response as demonstrated in Fig.~\ref{FigSCF} (d). The plasmonic
component visible through a divergent conductance $g_{sd}(\omega)$ with SC
vanishes for the non-SC calculation. This drastic change in the response
from the (divergent) plasmon-dominated to single-particle characteristics
is again clearly reflected in the ac charge density $\rho(z,\omega)$
calculated with (SC) and without (Non-SC) feedback shown in
Figs.~\ref{FigSCF} (e) and (f).
In the SC case, the charge density has a large amplitude at resonance with a
peak in the middle of the channel, a feature that is absent in the non-SC calculation.

\section{CONCLUSIONS}
We develop an approach for ac quantum transport within the nonequilibrium
Green function formalism, which allows to determine the frequency-dependent
charge and potential under excitation at a non-transport terminal within a
fully self-consistent framework.

The capability of our approach to determine the high-frequency properties of
systems in complex environments is demonstrated using a nanotube transistor
with an ac signal applied at its gate terminal. In the off state, the
dynamic conductance shows oscillations that originate from single-particle
excitations between quantized energy levels. When the device is operated in
the on state, the dynamic conductance exhibits discrete divergent peaks at
terahertz frequencies. These peaks are associated with plasmonic excitations
of the charge density at the resonant frequencies of the transistor acting 
as a quantum cavity. It is shown that the self-consistent coupling between
charge and potential is an essential component in the ac transport theory to
capture plasmon excitations of the system. A non-self-consistent approach
misses this important physics, and can only provide information about the
single-particle excitation spectrum.

The proposed approach is not limited to study the ac response of nanotube
devices, but can be applied to explore nonequilibrium, time-dependent
electronic and optical processes in other low-dimensional materials such
as nanowires, graphene, or molecules, including the exploration of their
collective excitation modes for novel plasmon-based nanoscale devices.

\section{ACKNOWLEDGMENT}
It is a pleasure to acknowledge discussions with Mark Lee, Clark
Highstrete, and Eric Shaner. This work was supported by the Laboratory
Directed Research and Development program at Sandia National Laboratories.
Sandia is a multiprogram laboratory operated by Sandia Corporation, a
Lockheed Martin Co., for the United States Department of Energy under
Contract No. DEAC01-94-AL85000. M.V. was supported at the
University of Alberta by the Natural Sciences and Engineering Research
Council (NSERC) of Canada.

\begin{appendix}
\section{DERIVATION OF THE NONEQUILIBRIUM PARTICLE DENSITY $G^{<}$}
In the following we detail the derivation for the particle density
$G^{<}(E,E')$, cf. Eq.~(\ref{EqGL}).
We start from the expression for the (time-domain) Dyson equation for
$G^{<}$ mapped onto the real axis utilizing Langreth
rules,\cite{Haug1998,LangrethPRB72} and symbolically written as
\begin{eqnarray}\label{AppEq:GL1}
G^{<} = \left( G_{0} + G_{0} U G\right)^{<} 
= G_{0}^{<} + G_{0}^{<} U G^{a} + G_{0}^{r} U G^{<}
\end{eqnarray}
where we have used that $U^{<}=U(\tau)\delta^{<}(\tau-\tau ') = 0$
for a time-local potential.\cite{Haug1998}
This equation can be re-arranged by collecting the $G_{0}^{<}$ terms first 
\begin{eqnarray}\label{AppEq:GL2}
G^{<} = G_{0}^{<} \left( 1 + U G^{a} \right) + G_{0}^{r}UG^{<}~. 
\end{eqnarray}
Equation~(\ref{AppEq:GL2}) can be solved through iteration by inserting the
expression for $G^{<}$ on the l.h.s. into the second term on the r.h.s., and
collecting now the $G_{0}^{<}\left( 1 + U G^{a} \right)$ elements.
After the first iteration one obtains
\begin{eqnarray}\label{AppEq:GL3}
G^{<} = \left( 1 + G_{0}^{r} U \right) G_{0}^{<} \left( 1 + U G^{a} \right)
+ G_{0}^{r} U G_{0}^{r} U G^{<}~.
\end{eqnarray}
The retarded Green function $G_{0}^{r}$ in the prefactor $(1 + G_{0}^{r} U)$ is 
the first term in Dyson's series, $G^{r} = G_{0}^{r} + G_{0}^{r} U G^{r}$, which
becomes more transparent when iterating one more time
\begin{eqnarray}\label{AppEq:GL4}
G^{<} &=&
\left(1 + G_{0}^{r} U + G_{0}^{r} U G_{0}^{r} U \right) G_{0}^{<} \left( 1 + U G^{a} \right) \\
&+& G_{0}^{r} U G_{0}^{r} U G_{0}^{r} U G^{<} \nonumber \\
&=& \left[ 1 + \left( G_{0}^{r} + G_{0}^{r} U G_{0}^{r} \right) U \right] G_{0}^{<}
\left( 1 + U G^{a} \right) \\
&+& G_{0}^{r} U G_{0}^{r} U G_{0}^{r} U G^{<}~.
\end{eqnarray}
Iterating to infinite order, this Dyson series converges towards $G^{r}$,
so that the final expression for the non-equilibrium particle density reads
\begin{eqnarray}\label{AppEq:GL5}
G^{<} = \left( 1 + G^{r} U \right) G_{0}^{<} \left( 1 + U G^{a} \right)~.
\end{eqnarray}
Equation~(\ref{EqGL}) in section II.B corresponds to Eq.~(\ref{AppEq:GL5})
after Fourier transform.

\section{Small Bias Expressions for $G^{<}$ and $\Sigma_{c}^{<}$}
The conductance associated with the particle and displacement current are response 
functions which relate the terminal current $I_{\alpha}$ with the terminal 
voltage $v_{\beta}$ in a {\it linear} manner. In order to derive a formula for the
conductance $g_{\alpha\beta}^{p}$ given in section III, one
needs linearized expressions for $\Sigma^{<}$ and $G^{<}$. 
These can be easily obtained from the Taylor expansion of the Fermi function
$f_{\beta} \equiv f_{\beta}(E) = 
1/\left[ 1 + e^{(E - \mu_{\beta,0} + e v_{\beta}) / k_B T} \right]$
to first order in the terminal voltages $v_{\beta}$, \textit{i.e.}
\begin{eqnarray}\label{AppEq:FermiLR}
f_{\beta} \approx f_{\beta,0} + e \tilde{f}_{\beta}~v_{\beta}~,~
\tilde{f}_{\beta} = -\frac{1}{k_{B}T}~f_{\beta,0} \left( 1 - f_{\beta,0} \right)~,
\end{eqnarray}
where $\mu_{\beta,0}$ is the chemical potential of terminal $\beta$ at zero bias,
and $f_{\beta,0} = 1 / \left[ 1 + e^{(E - \mu_{\beta,0})/k_B T} \right]$
the corresponding Fermi function. Inserting Eq.~(\ref{AppEq:FermiLR}) into
$\Sigma^{<}$ and $G^{<}$ one obtains the following set of linearized expressions
\begin{eqnarray}\label{AppEq:Sigma}
\Sigma_{\beta}^{<} &=& if_{\beta}\Gamma_{\beta} 
= \Sigma_{\beta,0}^{<} + \widetilde{\Sigma}_{\beta}^{<}~v_{\beta} \\
\Sigma_{\beta,0}^{<} &=& i\Gamma_{\beta} f_{\beta,0}~~,~~
\widetilde{\Sigma}_{\beta}^{<} = i\Gamma_{\beta}\tilde{f}_{\beta}
\end{eqnarray}
and
\begin{eqnarray}\label{AppEq:GL}
G_{0}^{<} &=& \sum_{\beta=s,d} G_{0}^{r} if_{\beta}\Gamma_{\beta} G_{0}^{a} 
= \bar{G}_{0}^{<} + \sum_{\beta=s,d} \widetilde{G}_{0,\beta}^{<}~v_{\beta} \\
\bar{G}_{0}^{<} &=& \sum_{\beta=s,d} G_{0}^{r}\Sigma_{\beta,0}^{<}G_{0}^{a}~~,~~
\widetilde{G}_{0,\beta}^{<} = G_{0}^{r}\widetilde{\Sigma}_{\beta}^{<} G_{0}^{a}~,
\end{eqnarray}
with $\Gamma_{\beta}=i\left( \Sigma_{\beta}^{r} - \Sigma_{\beta}^{a} \right)$
the broadening function.
\end{appendix}

\end{document}